# The least-used key selection method for information retrieval in large-scale Cloud-based service repositories

Jiayan Gu, University of Leicester, Leicester, UK, email: jg491@leicester.ac.uk, ORCID: 0000-0001-9355-5395
Ashiq Anjum*, University of Leicester, Leicester, UK (*Corresponding author), email: aa1180@leicester.ac.uk
Yan Wu*, Jiangsu University, Jiangsu, China (*Corresponding author), email: wuyan04418@ujs.edu.cn
Lu Liu, University of Leicester, Leicester, UK, email: l.liu@leicester.ac.uk
John Panneerselvam, University of Leicester, Leicester, UK, email: j.panneerselvam@leicester.ac.uk
Yao Lu, University of Leicester, Leicester, UK, email: yl604@leicester.ac.uk
Bo Yuan, University of Leicester, Leicester, UK, email: b.yuan@leicester.ac.uk

**Abstract**   As the number of devices connected to the Internet of Things (IoT) increases significantly, it leads to an exponential growth in the number of services that need to be processed and stored in the large-scale Cloud-based service repositories. An efficient service indexing model is critical for service retrieval and management of large-scale Cloud-based service repositories. The multilevel index model is the state-of-art service indexing model in recent years to improve service discovery and combination. This paper aims to optimize the model to consider the impact of unequal appearing probability of service retrieval request parameters and service input parameters on service retrieval and service addition operations. The least-used key selection method has been proposed to narrow the search scope of service retrieval and reduce its time. The experimental results show that the proposed least-used key selection method improves the service retrieval efficiency significantly compared with the designated key selection method in the case of the unequal appearing probability of parameters in service retrieval requests under three indexing models.

**Key words**   Cloud Computing; service computing; service retrieval; service addition; multilevel index model

## 1. Introduction

The rapid development of the Internet of Things (IoT) in recent years has led to the deployment and use of various applications of a distributed nature that generate huge amounts of data [1]. Cloud computing has been proposed to efficiently store and process large amounts of data and provide various services and resources according to user needs. Such as Amazon Web Services, Google App Engine and Microsoft Azure are already providing a variety of services to users with the help of cloud platforms.

With a vast number of services being hosted on the cloud, more and more researchers provide effective methods for service discovery and composition [2,3]. The inverted index [4] is the indexing model currently used for service retrieval in consistent repositories. However, the Inverted index has redundancy and is time-consuming which is not suitable for a large-scale service repository. In order to address this problem Wu et al. proposed a multilevel index model [5,6] to address the above issues. The efficiency of service retrieval is improved by eliminating redundancy, thus ensuring a reduced time for service discovery and composition. Fig. 1 shows the architecture of the multilevel index model, the core of which is used to store the services, containing the input and output parameters of the services and four levels of indexing for redundancy reduction (described in section 3.2). The service retrieval function takes a set of parameters as input and returns a set of services that can be invoked. The service discovery and composition system can quickly retrieve services from the service repository via the service retrieval API. In addition, the multilevel index model serves as an underlying storage structure for managing the services in the service repository, including the addition, deletion and replacement of services.

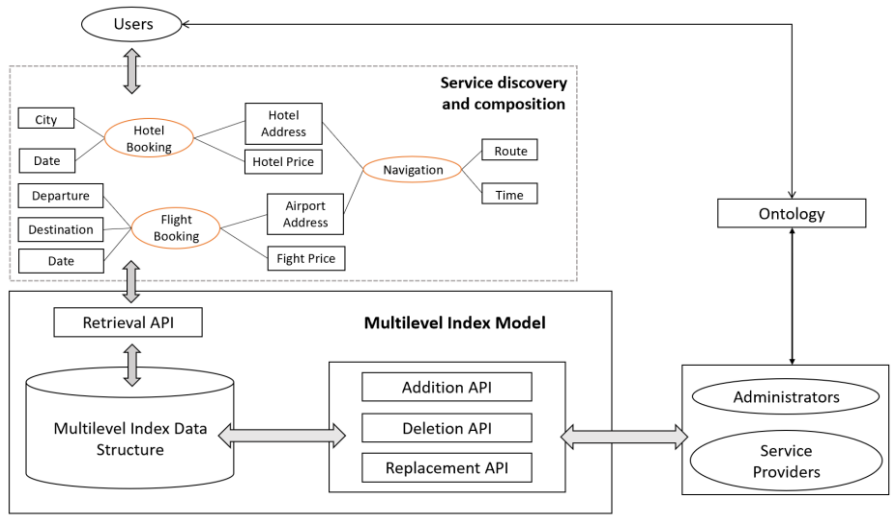

Fig. 1. An application scenario for the service multilevel index model.

The selection of keys is important for service indexing and retrieval in the multilevel index model. The original key selection method [6], the random key selection method [7], the maximum key count selection method [7] and the minimum key count selection method [7] have been proposed and evaluated as methods for selecting the "keys" of retrieval parameters under the assumption of equal probability of service parameter distributions. However, this assumption does not reflect the real situation of service parameter distribution, as some services inevitably have similar input or output parameters, or some services are frequently invoked by users resulting in unequal retrieval request parameters for each service. Large classes of services indexed by popular "keys" could slow down service retrieval process.

With this in mind, this paper proposes a novel least-used key selection method to enhance the multilevel index model for service retrieval and addition operations under the condition of unequal probabilities of service parameters. The main contributions of this paper are summarised as follows:

1. A least-used key selection method has been proposed to improve the efficiency of service retrieval under the condition of unequal probabilities of service parameters.

2. An enhanced multilevel index model under the unequal probability of service parameter distribution has been designed using the proposed least-used key selection method and compared with five existing key selection methods. Experimental evaluation demonstrates that our proposed least-used key selection method outperforms the other studied methods in terms of service retrieval efficiency.

The remainder of this paper is organised as follows: Section 2 presents a review of the recent related works about service discovery, composition and retrieval. Section 3 introduces the multilevel index model. Section 4 presents our proposed least-used key selection method to improve the efficiency of service retrieval under challenging conditions. Section 5 presents and discusses our experimental results. Section 6 concludes the paper along with outlining our future research directions.

**2. Related work**

*2.1 Service Discovery and Composition*

Service discovery technologies used till date are mainly based on service description language [8] such as XML, WSMO and OWL-S. [9] proposed a new semantic-aware web service discovery method, which was designed to provide relevant web services based on user queries. In addition, Bharti and Jindal [10] proposed a new search-based clustering strategy based on the heterogeneity of smart IoT devices and smart web services for service discovery methods in IoT environments. Although these methods are simple, their accuracy and recall rates are low, and there are still imperfections. The goal of service composition is to improve the reusability and utilisation of basic services. Yu et al. [11] proposed a new paradigm for automatic composition of Web services, adding keyword queries to the traditional graph search method based on input-output matching. Saleem et al. [12] proposed a service hierarchy model based on the awareness theory, which applies machine learning algorithms to learn the original Web Service selection scheme, so as to realise Web Service composition. However, these methods still have some drawbacks, such as low user satisfaction and high complexity. Huang et al. [13] stated that the number of dynamic services in a dynamic service network should be considered as the threshold value for evaluation, and further postulated to collect and map all the services as directed acyclic graphs and inverted index tables. Inverted indexing can reduce the service composition time, but incurs redundancy during the service retrieval process. Therefore, the development of an efficient service index model should effectively solve the aforementioned problems, and improve the efficiency of service discovery and composition through effective management and by reducing the scope of service retrieval.

*2.2 Service Retrieval*

The purpose of service retrieval is to find all services in the service repository that satisfy the user's needs as fast as possible, while excluding services of little relevance from the returned results.

The most common service retrieval methods are classified as Concept-based retrieval; Structure-based retrieval; Logical, inference-based retrieval; and Name-based retrieval, such as Sequential index [14], Tree-based index [15], Inverted index [4], and Hash table [16]. However, all these methods have some problems, such as Concept-based retrieval requires an excessive amount of upfront work in building the service concept ontology, and the accuracy of the ontology will directly affect the retrieval results [17]; Structure-based retrieval imposes an additional burden on the service provider when publishing the service [18]; Logical, Inference-Based retrieval, requires the support of a backend rule base, which needs to be built manually. In addition, the reasoning is implemented on the basis of building ontologies to achieve matching of retrieval requirements, which improves accuracy while also affecting retrieval efficiency [19,20], and Name-based retrieval is done by searching for keywords. However, it suffers from high complexity, low reliability and lower user satisfaction. Although methods such as the Tree-based index model and the Inverted index model can be used to narrow down the search space, these models come at the cost of service redundancy, which can increase the time for service discovery and composition.

Wu et al. proposed a multilevel index model to index services to reduce service retrieval redundancies and improve retrieval performance. The "key" is an important concept in the multilevel index model. One of the service input parameters is selected as the "key" for the service to be indexed. Wu *et al*. [6] proposed the original key selection method to select a key for a newly added service. Wu *et al*. [21] studied the effects of key selection methods to service retrieval. Kuang et al. [7] studied the key selection method in the multilevel index model, and introduced three different key selection methods including the minimum key count selection method, maximum key count selection method and random key selection method. Experiments have indicated that the random key selection method improves service addition operation. Gu et al. [22] studied the reason for service addition improvidence using the random key selection, and proposed a new key selection method, called the designated key selection method that can further reduce the service addition time without compromising the service retrieval efficiency and stability. However, these key selection methods do not consider the services parameters distribution under unequal probabilities in multilevel index models, which contradicts with the real-world situations where some services have the same and lapped parameters or some popular services parameters invoked more frequently than unpopular ones.

## 3. Multilevel Index Models

*3.1 The basic definition scenario*

The following definitions related with service are defined by Wu et al. [5,6].

**Definition 1.** A service is a composite $s= \{•s, s•, O\}$, where $•s$ is the set of input parameters, and $s•$ is the set of output parameters, and $O$ is a set of service attributes, e.g., *QoS*.

**Definition 2.** A user's request can be represented as $Q= (Q_p, Q_r)$, where $Q_p$ and $Q_r$ represent the set of service parameters provided by users and the set of service parameters requested by users, respectively.

**Definition 3.** Service retrieval can be defined as $Re(A, S) = \{s | •s \subseteq A \land s \in S\}$, where $A$ is the given parameter set and $S$ is a service set. The service retrieval parameter is often used to receive a set of parameters to define the user requirements, which usually returns the services invoked by the received set of parameters.

**Definition 4.** Service discovery Dc $(Q, L(O), S) = \{ s | Re(Q_p, S) \wedge Q_r \subseteq s\bullet \wedge s.O \angle L(O) \}$, where $L(O)$ is a set of constraints for any other attributes, $S$ is a service set, and $s.O \angle L(O)$ means that $s$ satisfies these constraints.

In this paper, service retrieval is defined to find services that can be invoked according to users' provided parameter sets. Service discovery is to find services that can be invoked and satisfy users' requirements according to their requests. As shown in Definition 4, service retrieval is a part of service discovery. If the time for service retrieval is reduced, then the time for service discovery will also be reduced. Service composition requires successive service retrievals to combine different services to satisfy users' requirements that any single service cannot meet complete the users' requests. Therefore, efficient service retrieval improves the efficiency of service discovery and composition.

The same three services from the service discovery and combination scenario in Fig. 1 are used as an example, namely 'hotel booking', 'flight booking' and 'navigation', all of which are indexed in the service repository. Fig. 2 illustrates the redundancy in the classic inverted index and illustrates the importance of the key selection method by comparing Figs. 2 to 4.

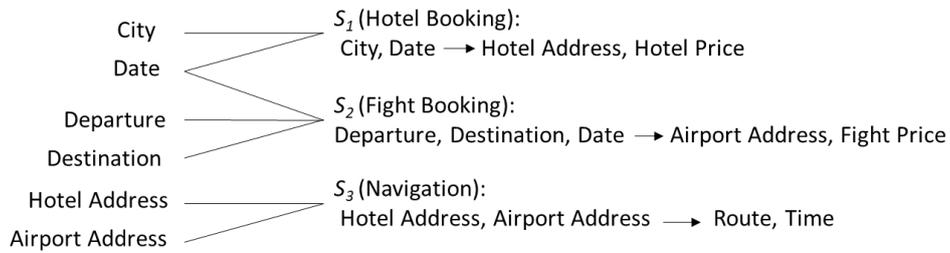

Fig. 2. The inverted index of three services used in Fig. 1.

Fig. 2 shows the inverted index of these three services. Assume *city* and *date* are given for a retrieval of $s_1$, then, from *city*, $s_1$ is searched, and from *date*, $s_1$ and $s_2$ are searched. The services have been searched three times in total ($s_1$ is searched once and $s_2$ is searched twice) before $s_1$ can be retrieved.

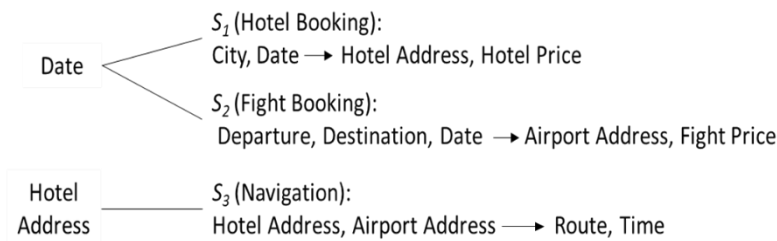

Fig. 3. The key index of three services used in Fig. 1.

Fig. 3 illustrates the idea of the key index. *Date* is the key of $s_1$ and $s_2$, and *hotel address* is the key of $s_3$. For the same retrieval of $s_1$, from *city*, no service needed to be searched since *city* is not a key, and from *date*, $s_1$ and $s_2$ are searched. The services have been searched twice in total ($s_1$ is searched once and $s_2$ is searched once) before $s_1$ can be retrieved, which is less than that of the inverted index (i.e. three times). Fig. 3 illustrates how the key reduces redundancy and further improves service retrieval performance.

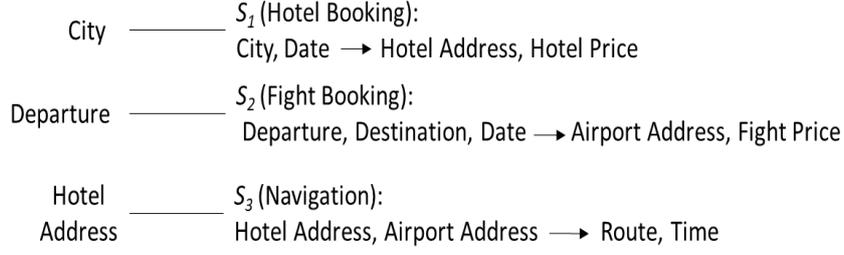

Fig. 4. A different key index of three service used in Fig. 1, which is used to illustrate that the key selection affects retrieval efficiency.

Fig. 4 shows a different key index. *City*, *departure* and *hotel address* are the keys of $s_1$, $s_2$ and $s_3$, respectively. For the same retrieval of $s_1$, from *city*, $s_1$ is searched, and from *date*, no service needed to be searched since *date* is not a key. Hence, the targeted $s_1$ is searched only once, which is less than that of the index illustrated in Fig. 3 (twice). From the example illustrated in Figs. 3 and 4, it can be seen that the key selection methods can affect service retrieval performance. This paper focuses on the study of the key selection method under the condition of unequal service invoking frequency.

*3.2 Multilevel Index Models*

Based on the characteristics of integrity, non-redundancy and certainty of equivalence relations, Wu et al. [23] proposed an efficient multilevel index model for service retrieval based on equivalence relations, which stores and manages large-scale service repositories. This model can reduce the scope of service retrieval quickly and can improve the efficiency of the service retrieval process, thus the time for service discovery and service composition can be reduced. The multilevel index model is divided into four levels, which are:

- The First Level Index *($L_1I$)*: This is an index between a service $s$ and a similar class $C_s$ if $s \in C_s$.

- The Second Level Index *($L_2I$)*: This is an index between a similar class $C_s$ and an input-similar class $\mathcal{C}_{is}$ if $C_s \in \mathcal{C}_{is}$.

- The Third Level Index *($L_3I$)*: This is an index between an input-similar class $\mathcal{C}_{is}$ and a key class $\mathcal{C}_k$ if $\mathcal{C}_{is} \in \mathcal{C}_k$.

- The Fourth Level Index *($L_4I$)*: This is an index between a key class $\mathcal{C}_k$ and a *key*, $key \in K$ if $f_k(\mathcal{C}_k) = key$.

The relationship diagram of the entire multilevel index model is shown in Fig. 5. Firstly, the service set $S$ is divided into many subsets, and each subset contains the same input and output parameters, which are called similar classes and denoted as $C_s$. Therefore, the index between service $S$ and $C_s$ is denoted as $L_1I$, which can reduce the redundancy of repeated retrieval caused by services having the same input and output parameters. Secondly, the services containing the same input in the similar class $C_s$ are divided into a class, called input-similar class and denoted as $\mathcal{C}_{is}$. The index between $C_s$ and $\mathcal{C}_{is}$ is denoted as $L_2I$, which can reduce the redundancy caused by the same input parameters of services. Then, the services that have the same key in the similar class $\mathcal{C}_{is}$ are divided into a set, which is called key class, denoted as $\mathcal{C}_k$, while the index between $\mathcal{C}_{is}$ and $\mathcal{C}_k$ is denoted as $L_3I$. The unique index established between each $\mathcal{C}_k$ and key value $K$ is denoted as $L_4I$, which can improve the service retrieval efficiency by selecting a unique key $K$ to retrieve the required services.

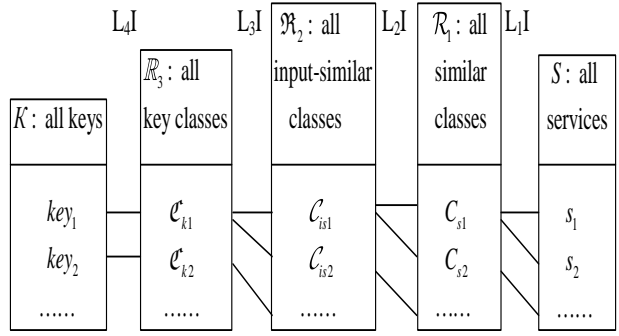

Fig. 5. The multilevel index model (full index model) of services.

Fig. 6 shows a specific multilevel index of services. There are five services $s_1$-$s_5$ in the service repository. Firstly, $s_1$ and $s_2$ compose a similar class since they have the same inputs and outputs. Other services compose different similar classes, respectively. Secondly, the first and the second similar class compose an input-similar class since they have the same inputs. Other similar classes compose different input-similar classes, respectively. Finally, the second and the third input-similar classes compose a key class since they have the same key. The other input-similar class composes a key class alone.

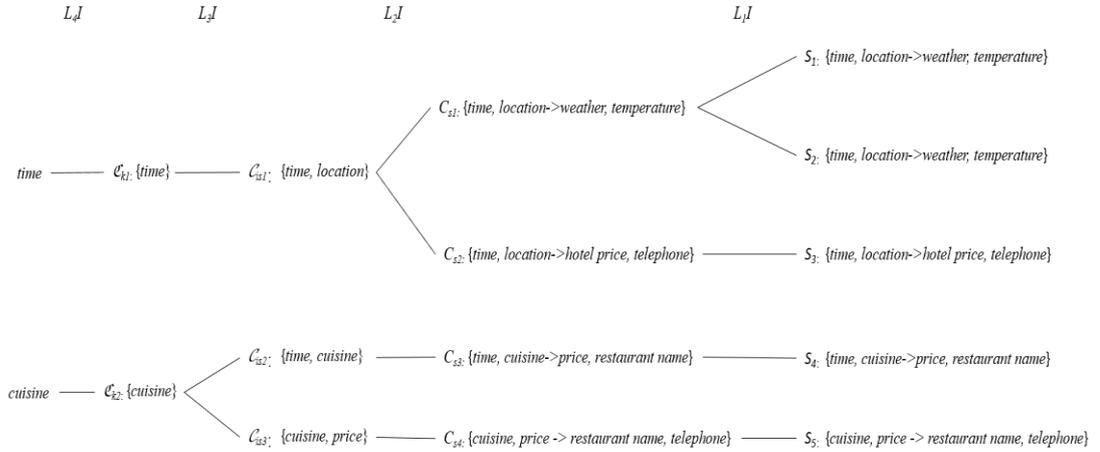

Fig. 6. An example of the multilevel index.

*3.3 Flexible Deployment*

The multilevel index model can be deployed using three different methods [21] including the primary index model ($L_3I$-$L_4I$), the partial index model ($L_2I$-$L_4I$) and the multilevel index model ($L_1I$-$L_4I$). Both the partial and primary index models, as shown in Fig. 7 and Fig. 8, can be used for different service repositories with different sizes and characteristics.

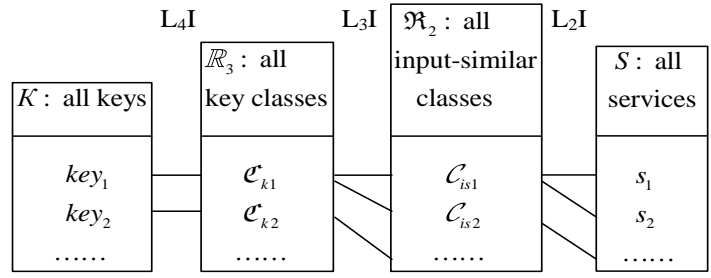

Fig. 7. The partial index model of services.

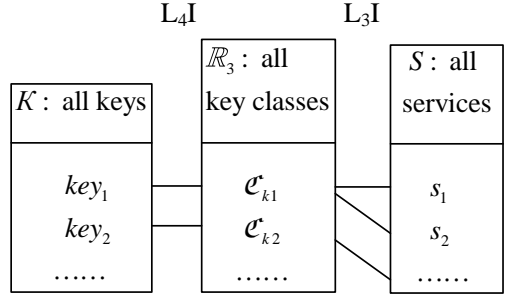

Fig. 8. The primary index model of services.

*3.4 Different key selection methods*

The following five key selection methods have been studied and evaluate our proposed key selection method against their efficiency in terms of service retrieval and addition operation.

1) The original key selection method

The original key selection method, which makes $|\mathcal{C}_k|$ as close to $\sqrt{|R_2|}$ as possible. **Algorithm** 1 presents the operation of the original key selection method.

---
*Algorithm* 1. Original key selection method
---
Input: *s*
Output: key of *s*
1. Try to find an input similar class that has the same inputs with *s*.
2. If an input similar class is found, select its key as the key of *s* and return the key.
3. Try to find a set of input similar classes such that their keys are contained in the inputs of *s* and the size of any similar class is less than $\sqrt{|R_2|}$.
4. If the input similar class set is empty, randomly select an input of *s* as its key and return the key.
5. Find an input similar class with the biggest size, select its key as the key of *s* and return the key.
---

2) The minimum key count selection method

The principle of the minimum key count selection method uses an existing key as the key of newly added services and maintains key classes as smaller as possible. If the parameters of a given service cannot be found in the existed key classes, then randomly select an input of *s* as its key.

---

*Algorithm* 2. Minimum key count selection method

---

Input: *s*

Output: a key

1. Try to find an input similar class that has the same inputs with *s*.
2. If the input similar class is found, select its key as the key of *s* and return the key.
3. Try to find an input similar class such that its key is contained in the inputs of *s*.
4. If an input similar class is found, select its key as the key of *s* and return the key.
5. Randomly select an input of *s* as its key and return the key.

---

3) The maximum key count selection method

On the contrary to the minimum key count selection method, the maximum key count selection method uses the existing key as the key of newly added services and maintains the number of key classes as bigger as possible.

---

*Algorithm* 3. Maximum key count selection method

---

Input: *s*

Output: key of *s*

1. Try to find an input similar class that has the same inputs with *s*.
2. If the input similar class is found, select its key as the key of *s* and return the key.
3. Try to find an input of *s* such that no input similar class uses the input as a key.
4. If such an input of *s* is found, select the input as the key of s and return the key.
5. Randomly select an input of *s* as its key and return the key.

---

4) The random key selection method

The random key selection method randomly selects one of the service input parameters as the key of the service through a random number function.

---

*Algorithm* 4. Random key selection method

---

Input: *s*

Output: key of *s*

1. Try to find an input similar class that has the same inputs with *s*.
2. If the input similar class is found, select its key as the key of *s* and return the key.
3. Randomly select an input of *s* as its key and return the key.

---

5) The designated key selection method

The designated key selection method narrows down search space when a new service is added to the partial or full index, and determines a unique parameter in the service input parameter as the key.

---

*Algorithm* 5. Designated key selection method

---

Input: *s*

Output: key of *s*

1. $sum=0$;

2. For each input parameter α of *s*
3. {*sum=sum+*α.*id*;}
4. *i=sum* mod *c*; // (*c* denotes the count of the input parameter of *s*.)
5. select the $i^{th}$ input of *s* as its key and return the key.

---

## 4. A Least-used key selection method

In a real-world scenario, some services have the same and lapped parameters or some popular services parameters invoked more frequently, while others show the opposite trend and are rarely invocated. The key selection methods of the multilevel index model proposed in [7,22,6] do not consider the unequal distribution of service parameters, resulting in the test results of service retrieval and addition under the multilevel index model not conforming to the actual situation. In order to deal with this situation more efficiently, we propose a novel key selection method, called the least-used key selection method, which can further improve the service retrieval efficiency in the index model.

Before designing the key selection method, an enhanced multilevel index model usually corresponds to two situations that result in unequal service parameters. One is the unequal probability of parameters appearing in the service retrieval request sets, and the other is the unequal probability of parameters appearing in the service inputs. Both the equal and unequal probability of parameters in the multilevel index models do not affect the service retrieval efficiency [7]. This implies that parameter probabilities of service inputs do not impact the retrieval efficiency to any significant level, due to the existence of service input parameters in the multilevel index. On the contrary, service request parameters given by users are outside the multilevel index, thus their distribution significantly affects the retrieval performance. For this reason, a least-used key selection method is proposed to choose appropriate keys according to appearing probabilities of parameters appearing in retrieval request sets.

The proposed least-used key selection method is based on the following hypothesis: If a service is frequently invoked, its corresponding key class should be relatively smaller; on the contrary, if a service is rarely invoked, its corresponding key class should be relatively larger.

The proposed method is first illustrated intuitively. According to **Definition 3** discussed previously, in a service retrieval request *Re* (*A*, *S*), *A* is a set of requests submitted by the user and *S* represents a collection of services. Suppose a request set {{*a*, *b*}, {*b*, *c*}, {*c*, *a*}}, where the parameters *a*, *b*, and *c* are invocated with an even probability. However, in a real scenario, every parameter to characterise an even probability is nearly impossible. For example, in a request set {{*a*, *b*}, {*a*, *c*}, {*a*}}, *a* appears more frequently than other parameters. If *a* is a key, then more services are retrieved. The least-used key selection method proposed in this paper avoids the need to select a as a key.

Next, the method was proven to be correct. Let $x_i=|\mathcal{C}_{ki}|$, i.e., the total number of input-similar classes contained in $\mathcal{C}_{ki}$; and $m=|\mathfrak{R}_2|$, i.e., the total number of all input-similar classes. Therefore, the following formula (1) can be obtained.

$$x_1 + x_2 + \cdots + x_n = m, \quad (0 \leq x_i \leq m). \qquad (1)$$

Let $p_i$ denote the retrieval probability for $\mathcal{C}_k$, and $y$ denote the total number of input-similar classes being retrieved. Then, the following formula (2) can be obtained.

$$y = p_1 x_1 + p_2 x_2 + \cdots + p_n x_n, \quad (\sum_i p_i = 1) \quad (2)$$

The optimal target is to maintain $y$ as small as possible. According to rearrangement inequality [24] (also known as sequence inequality), if $x_1 \geq x_2 \geq \ldots \geq x_n$ and $p_1 \leq p_2 \leq \ldots \leq p_n$, then $y$ (called as reversed sum) is minimised. Generally, every $p_i$ value is known in a real-world situation. Therefore, the proposed least-used key selection method finds the most appropriate key for a newly added service and efficiently minimizes $y$.

According to the above analysis, the proposed least-used key selection algorithm is illustrated in *Algorithm* 6.

---

*Algorithm* 6. Least-used key selection method

---

Input: *s*
Output: key of *s*
1. Select the first input parameter of *s* as the key *k*
2. For each input parameter *a* of *s*
3.    If (*a. appearing_ probability* < *k. appearing_ probability*)
4.      {*k=a*}
5. Select *k* as the key of *s* and return *k*.

---

The *a. appearing_probability* denotes the probability of the parameter *a* appearing in a request set. Moreover, the method of distributing the parameters with unequal probability in the request set will be given in *Algorithm* 7 below. The objectives of *Algorithm* 6 are to check each parameter of a newly added service and to determine its key characterising the smallest appearing probability. In this way, the value of $y$ in equation 2 is minimised to its lowest level. Hence, as fewer input-classes as possible are searched during the retrieval operation.

In the same way as above, the case where the service input parameters are based on unequal probability distributions should also be considered. The proposed least-used key selection method is evaluated under the scenarios of unequal probabilities of parameters appearing in service inputs, despite the fact that such unequal probabilities are not known to affect the service retrieval and addition efficiencies. Hence, step 3 in *Algorithm* 6 is modified as follows in order to evaluate its retrieval efficiency under the condition of unequal probabilities appearing in service inputs.

If (*a.appearing_ probability_in_service inputs* < *k.appearing_ probability_in_service inputs*)

## 5. Experimental results and analysis

*5.1 Experimental environment and settings*

Our simulation platform is developed in Microsoft Visual Studio using C#. Each component is built with low coupling capacity and can be modified and upgraded separately. In our experiments, the least-used key selection method and the other five key selection methods including the original key selection method, the random key selection method, the minimum key count selection method, the maximum key count selection method and the designated key selection method are evaluated respectively under the primary index model, partial index model and full index model with a different situation of services parameters distribution.

Our experimental approach includes the following steps: firstly, design an enhanced multilevel index model under unequal probability that a probability density function is selected to simulate the unequal appearing probabilities of parameters. Secondly, our least-used key selection method and the other five key selection methods are integrated into the primary, partial and full index models. Finally, the service retrieval and addition efficiencies of the six key selection methods are evaluated in the three index models under different parameter distribution conditions.

In the first step, the Monte Carlo method [25] is incorporated into our test platform to generate a selected distributed random number as service input parameter or service retrieval request parameter under an unequally appearing probability, as shown in *Algorithm* 7.

---
*Algorithm* 7. Monte Carlo random number generator using probability density function

---
Input: *minX, maxX, minY, maxY, f(x)*
Output: a random number following the probability density function
1. do
2. $x$ = Random (*minX, maxX,*); // Generate the valid random number between the *minX* and *maxX* values on the X-axis
3. $y$ = Random (*minY, maxY*); // Generate a valid random number between *minY* and *maxY* on the Y-axis
4. if ($y \leq f(x)$) return $x$;
5. While($y > f(x)$)

---

Our experiment uses the Monte Carlo method to generate random numbers as service input parameters or service retrieval request parameters under an unequally appearing probability, that is, two independent random variables through a suitable probability density function are used to generate random numbers that meet the requirements. The probability density function is as follows.

$$f(x) = l(x - q) + q, 0 \leq x < q \quad (3)$$

where, $l$ is the slope; and $q=|P|$, where $P$ is a set of all parameters. When $f(x)$ is substituted into *Algorithm* 7, *minX* = *minY* = 0, *maxX* = *maxY* = $q$. Different values of $x$ are represented to denote different parameters. In our test platform, $l$ can be set to different values for different unequal distributions of the service parameters. If $l \leq 0$, then the distribution becomes even.

*5.2 Experimental results analysis*

It was tested in a multilevel index model with 50,000 services and the size of all the parameter sets is set to 1000. Each service has 10 input and 10 output parameters. In addition, each retrieval request contains 32 parameters and each dataset contains 1000 retrieval requests. In order to compare the efficiencies of the key selection methods, 20 artificial data sets are used to test the efficiencies of the six key selection methods. Their experimental results are as follows.

Fig. 9 presents the retrieval time of the six key selection methods in the primary index model under an equal parameter appearing probability. From Fig. 9, there is no obvious distinctiveness about the retrieval time with reference to the six key selection methods. These results also verified our previous work that the key selection methods do not affect the retrieval efficiency to any noticeable level under equal probabilities of service invocations [7].

Unequal probabilities of parameters appearance in service inputs and service retrieval requests are tested respectively. Fig. 10 presents the retrieval time of the six key selection methods on primary index models, under an unequal appearing probability of parameters in service inputs. Similar to Fig. 9, the average service retrieval time of the six key selection methods remains very similar, as illustrated in Fig. 10. The results verified our previous work [7] that the size of the key set does not affect the service retrieval efficiency in the multilevel index models.

Fig. 11 illustrates the service retrieval time of the six key selection methods in the primary index models, under an unequal appearing probability of parameters in service retrieval requests. Service retrieval request parameters are generated by users outside the multilevel index models. Therefore, different key selection methods have different service retrieval efficiencies. The proposed least-used key selection method exhibits the best performance, while the maximum and minimum key count selection methods cost most time due to their key selection methods do not optimize the retrieval time of the services.

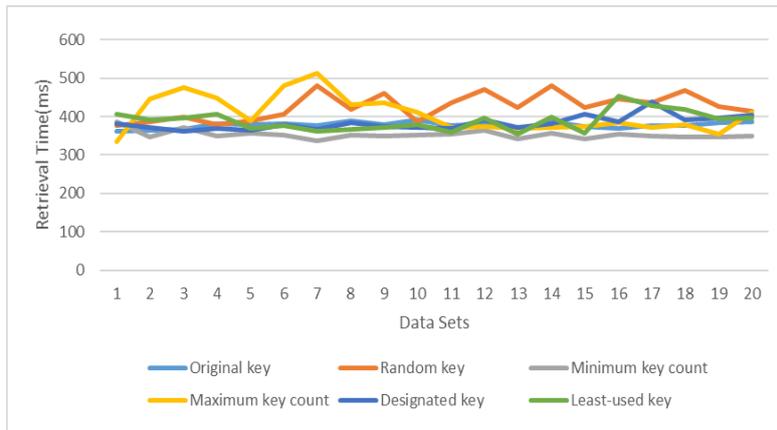

Fig. 9. Retrieval time on primary index models of the six key selection methods with equal appearing probabilities of parameters.

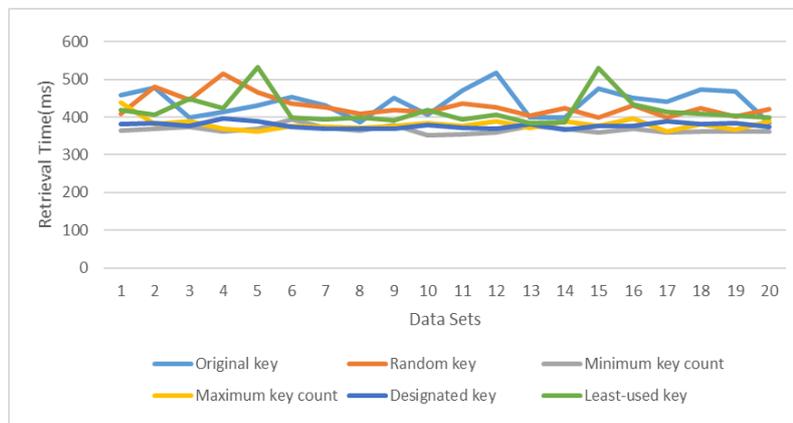

Fig. 10. Retrieval time on primary index models generated using the six different key selection methods with unequal appearing probabilities of parameters in services inputs.

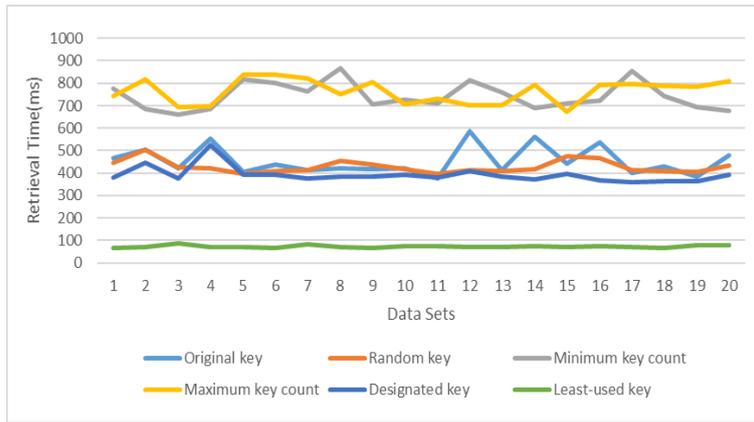

Fig. 11. Retrieval time on primary index models generated using the six different key selection methods with unequal appearing probabilities of parameters in services retrieval requests.

In Fig. 12, Fig. 13 and Fig. 14, service addition efficiencies of different key selection methods in the primary index model were tested under both equal and unequal appearing probabilities of services parameters, respectively. From these figures, the performance of the six key selection methods under each scenario has no obvious distinctiveness since the primary index does not retrieve input-similar classes for the service addition operation.

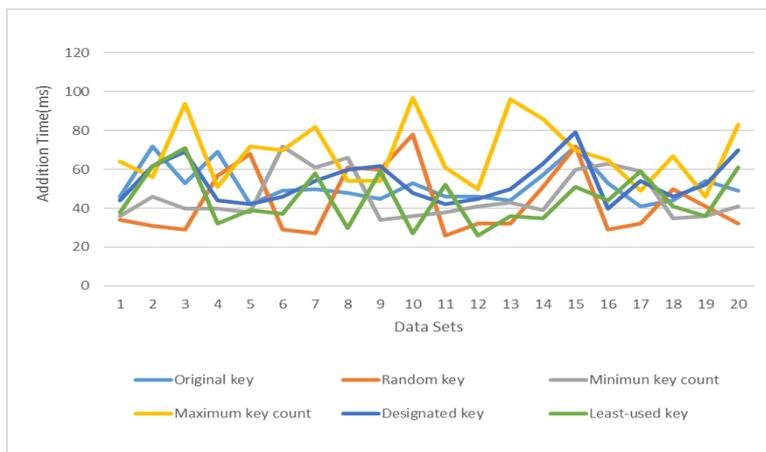

Fig. 12. Addition time on primary index models generated using the six key selection methods with equal appearing probabilities of parameters.

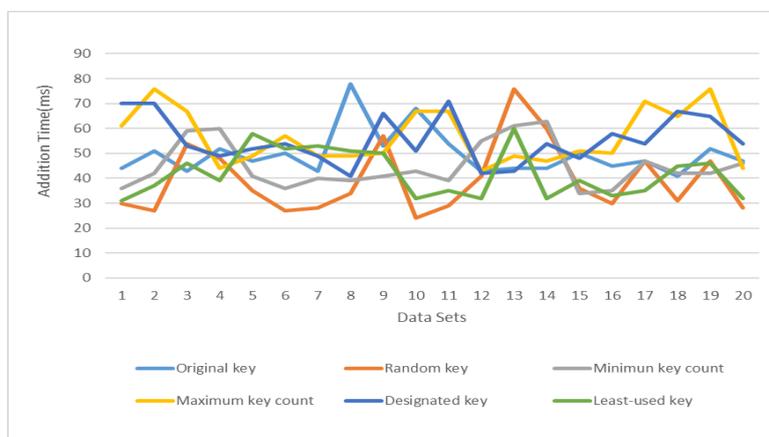

Fig. 13. Addition time on primary index models generated using the six key selection methods with unequal appearing probabilities of parameters of service inputs.

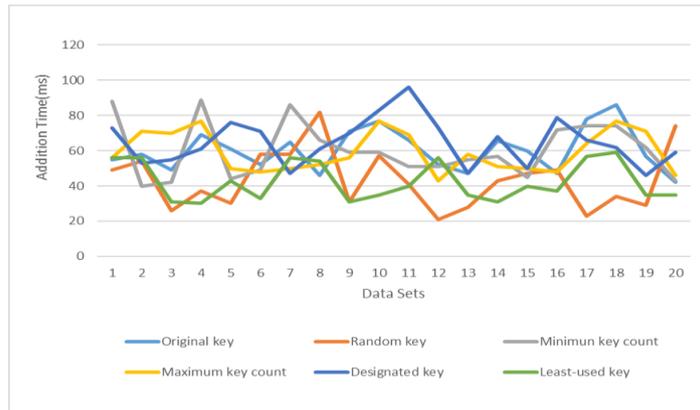

Fig. 14. Addition time on primary index models generated using the six key selection methods with unequal appearing probabilities of parameters in service retrieval requests.

Since the partial index and full index models are very similar except the fact that the partial index model is less time-consuming than the full index model, thus only the results of the full index model are exhibited. Retrieval and addition time with related to the six key selection methods in a full index model are very similar to that in a partial index model except the retrieval time is slightly longer. Therefore, only the experimental results in the full index model are shown. Fig. 15, Fig. 16 and Fig. 17 present the retrieval time of the six key selection methods in the full index models under both equal and unequal appearing probabilities of parameters appearing in service inputs and retrieval requests, respectively. The results are similar to the ones of the six key selection methods under primary indexing, and the least-used key selection method is still the best one that significantly reduces service retrieval time when the parameters with unequal appearing probability in the service retrieval requests.

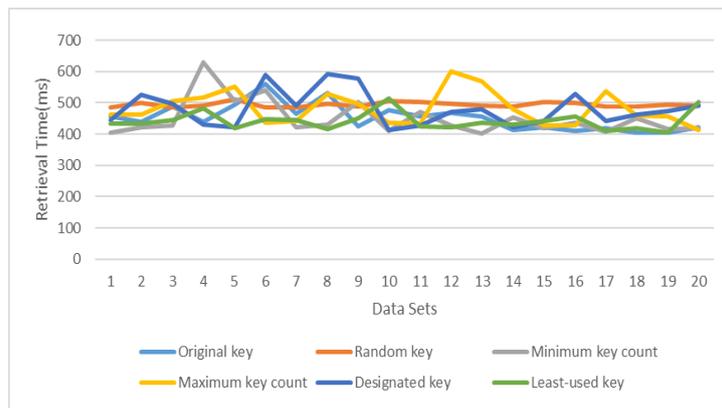

Fig. 15. Retrieval time on full index models generated using the six key selection methods with equal appearing probabilities of parameters.

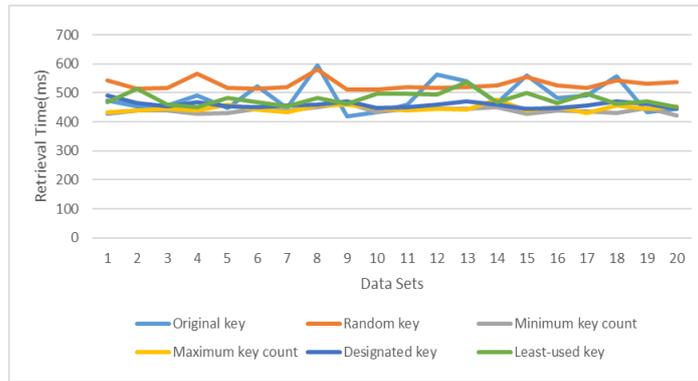

Fig. 16. Retrieval time on full index models generated using the six different key selection methods with unequal appearing probabilities of parameters in service inputs.

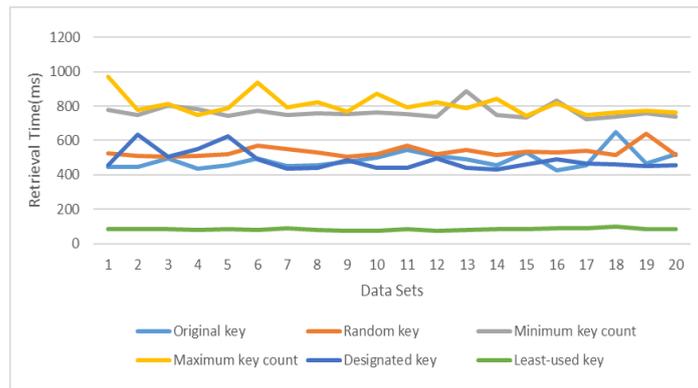

Fig. 17. Retrieval time on full index models generated using the six different key selection methods with unequal appearing probabilities of parameters in service retrieval requests.

In both the partial and full index models, when a new service is added, the original key selection method, max key count selection method, min key count selection method and the random key selection method, all require to retrieve a proper input-similar class containing the same input parameters with the new service. However, the designated key selection method and the proposed least-used key selection method do not need such a process, therefore they both have distinctive advantages for service addition over the other four methods. Fig. 18, Fig. 19 and Fig. 20 present the addition performances on full index models related to the six key selection methods under different parameter distribution conditions.

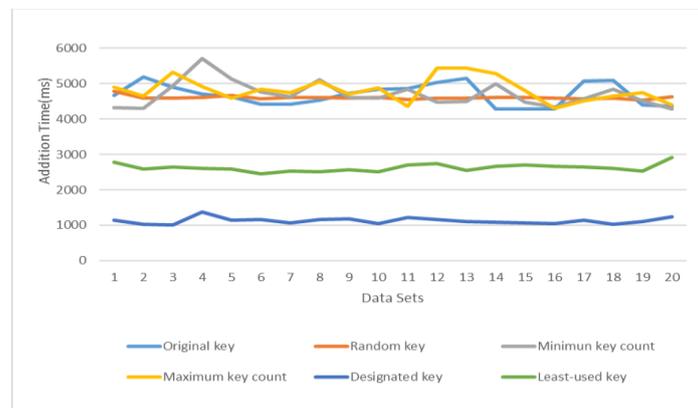

Fig. 18. Addition time on full index models generated using the six key selection methods with equal appearing probabilities of parameters.

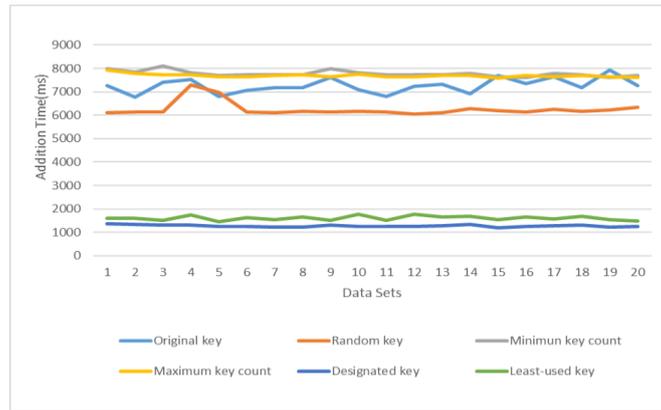

Fig. 19. Addition time on full index models generated using the six key selection methods with unequal appearing probabilities of parameters in service inputs.

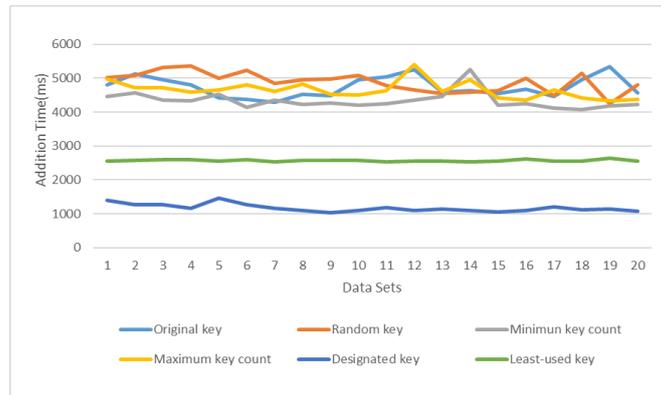

Fig. 20. Addition time on full index models generated using the six key selection methods with unequal appearing probabilities of parameters in service retrieval requests.

To summarise their strengths, the six methods are rated as 'fair', 'good' and 'excellent' by comparing the speed of service retrieval time and service addition time for the different key selection methods under different conditions. Since the results in Figs. 9, 10 and 12-16 do not have obvious distinctiveness, their results are rated as "average". In other test cases, average values of the results are used to rate them. The ratings for the different key selection methods in Figs. 11 and 17-20 are listed in Table 1. In order to exclude subjective interference, a clustering method is used to rate them. In recent years, spectral clustering has emerged as one of the most popular modern clustering algorithms. It is simple to implement, can be solved efficiently using standard linear algebra software, and frequently outperforms traditional clustering algorithms. In [26] introduced the family of spectral clustering algorithms, and compared to the "traditional algorithms" such as k-means or single linkage, spectral clustering has many fundamental advantages. Spectral clustering is a family of methods to find K clusters using a matrix's eigenvectors. One notable advantage of spectral clustering is its ability to cluster "points" that are not necessarily vectors, and to use for this a "similarity", which is less restrictive than a distance. The flexibility of spectral clustering is another advantage; it can find clusters of arbitrary shapes under realistic separations [27]. Since spectral clustering is highly adaptable to data distribution, it can cluster similar data into a similar space, in addition, the spectral clustering will be effective when the number of clustered categories is small. In this experiment, the categories are only divided into 3 classes, therefore, spectral clustering was selected to better meet the classification requirements. The final rating results are shown in Table 1. Overall, the min and max key count selection methods got the most "fair" ratings, the random and original key selection methods are within the moderate level, and the designated and least-used key selection methods divided all the "excellent" ratings.

In the case of the unequal probability of parameters appearing in service retrieval requests, the proposed least-used key selection method shows significant superiority in reducing service retrieval time no matter in primary, partial or full index models, where the least-used key selection method improves over 450% retrieval efficiency than the designated key selection method in these conditions. In contrast, the designated key selection method and the least-used key selection method both show significant superiority over other methods in adding services in all cases under the partial and full indexing models regardless of the service parameters distribution conditions. Compared with the least-used key selection method, the designated key selection method shows around 100% improvement in service adding efficiency under partial and full index models. Therefore, the least-used key selection method has an obvious advantage for service repositories with frequently retrieval requests, while the designated key selection method has an advantage for service repositories with frequently service addition and deletion operations.

Table 1. Retrieval and addition performances on primary/partial/full index models generated using the six key selection methods under different parameter distributions.

| Performance on primary index model | Original key | Random key | Min Key count | Max key count | Designated key | Least-used key |
|---|---|---|---|---|---|---|
| Retrieval<br>Equal parameter appearing probability | Average* | Average | Average | Average | Average | Average |
| Retrieval<br>Unequal appearing probability of service input parameters | Average | Average | Average | Average | Average | Average |
| Retrieval<br>Unequal appearing probability of service retrieval request parameters | Good<br>(453.9ms) | Good<br>(428.4ms) | Fair<br>(742.7ms) | Fair<br>(763.7ms) | Good<br>(392.0ms) | Excellent<br>(72.7ms) |
| Addition<br>Equal parameter appearing probability | Average | Average | Average | Average | Average | Average |
| Addition<br>Unequal appearing probability of service input parameters | Average | Average | Average | Average | Average | Average |
| Addition<br>Unequal appearing probability of service retrieval request parameters | Average | Average | Average | Average | Average | Average |
| **Performance on partial/full index model** | **Original key** | **Random key** | **Min Key count** | **Max key count** | **Designated key** | **Least-used key** |
| Retrieval<br>Equal parameter appearing probability | Average | Average | Average | Average | Average | Average |
| Retrieval<br>Unequal appearing probability of service input parameters | Average | Average | Average | Average | Average | Average |
| Retrieval<br>Unequal appearing probability of service retrieval request parameters | Good<br>(485.1ms) | Good<br>(533.3ms) | Fair<br>(765.2ms) | Fair<br>(807.7ms) | Good<br>(482.9ms) | Excellent<br>(84.2ms) |
| Addition<br>Equal parameter appearing probability | Fair<br>(4696.4ms) | Fair<br>(4607.9ms) | Fair<br>(4698.9ms) | Fair<br>(4828.9ms) | Excellent<br>(1122.2ms) | Good<br>(2626.9ms) |
| Addition<br>Unequal appearing probability of service input parameters | Fair<br>(7262.5ms) | Fair<br>(6264.5ms) | Fair<br>(7771.1ms) | Fair<br>(7694.0ms) | Excellent<br>(1271.6ms) | Good<br>(1606.5ms) |
| Addition<br>Unequal appearing probability of service retrieval request parameters | Fair<br>(4740.5ms) | Fair<br>(4887.0ms) | Fair<br>(4339.6ms) | Fair<br>(4652.4ms) | Excellent<br>(1170.4ms) | Good<br>(2567.6ms) |

*Average means all key selection methods have similar performance.

## 6. Conclusions and future directions

The existing key selection methods of the multilevel index model do not consider the effects of an unequal probability distribution of service parameters on service retrieval and addition performances. This paper proposed a new key selection method, namely the least-used key selection method and an enhanced multilevel index model has been designed to deal with these situations with higher performance. The performance of the proposed least-used key selection method is evaluated against five key selection methods under various conditions including equal probabilities of parameter distributions, and unequal probabilities of parameters distribution in service inputs and retrieval requests on the primary index, partial index and full index models, respectively. The experimental results show that the proposed least-used key selection method and the designated key selection method are superior to other methods, and the least-used key selection method is the best one for service retrieval.

In our experiments, the distributions of service parameters are known as the least-used key method. In the real-world, the distributions change from time to time. In our further work, we will study an adaptive key selection method based on the current work. We plan to evaluate and improve the performance of the proposed least-used key selection method under more complex and dynamic conditions, while further optimizing the service addition time.

**Declarations**

*Availability of data and materials*

Not applicable.

*Competing interests*

The authors declare that they have no competing interests.

*Funding*

Not applicable.

*Authors' contributions*

This research paper was co-authored by seven authors. Therefore, any author was involved in each part of the paper. However, the basic role of each author is summarized as follows: J.G. was the designer of the proposed model and methods and was responsible for the experiments of the proposed method with the support of A.A. and Y.W., L.L. assisted J.G. with the model design. J.P., B.Y. and Y.L. were the main reviewers of the paper, giving effective suggestions for improvement. All authors have read and agreed to the published version of the manuscript.

*Acknowledgements*

Not applicable.